\documentclass[
aps,
prd,
onecolumn,
superscriptaddress,
nofootinbib,
amsmath,
amssymb
]{revtex4-2}
\usepackage{float}
\usepackage[utf8]{inputenc}
\usepackage[T1]{fontenc}
\usepackage{graphicx}
\usepackage{tikz}
\usetikzlibrary{arrows.meta}
\usepackage{bm}
\usepackage{pstricks}
\usepackage{pst-node}
\usepackage[colorlinks=true,
linkcolor=blue,
citecolor=blue,
urlcolor=blue
]{hyperref}

\begin{document}
	\def\ba{\begin{eqnarray}}
		\def\ea{\end{eqnarray}}
	\def\w{\wedge}
	\def\d{\mbox{d}}
	\def\D{\mbox{D}}
	\def\w{\wedge}
	\def\d{\mbox{d}}
	\def\D{\mbox{D}}
	\newcommand{\tr}{\mathrm{Tr}}
	\newcommand{\hs}{\ast}
	\newcommand{\gm}{\bm{g}}

\title{On the Motion of a Point Electric Charge in the Field of\\ Wu-Yang Monopoles in Bertotti-Robinson Spacetimes}
\author{Tekin Dereli}
\email{tdereli@ku.edu.tr}
\affiliation{
	Department of Physics,
	Ko\c{c} University,
	34450 Sar{\i}yer, \.{I}stanbul, Turkey
}
\affiliation{Faculty of Aviation and Aeronautical Sciences,
	\"Ozy\u{e}\u{g}in University,
	34794 \c{C}ekmek\"{o}y, \.{I}stanbul, Turkey}
\author{Yorgo Senikoglu}
\email{ysenikoglu@gsu.edu.tr}
\affiliation{
	D\'{e}partement de Math\'{e}matiques,
	Universit\'{e} Galatasaray,
	34349 Be\c{s}ikta\c{s}, \.{I}stanbul, Turkey
}

\date{\today}
	
\begin{abstract}
	\noindent
	We study the $SU(2)$ Einstein-Yang-Mills system with non-minimal $RF^2$-type
	couplings of Yang-Mills fields to curvature, formulated with exterior differential forms
	and a Lagrange-multiplier constraint for the Levi-Civita connection. We show that
	the conformally flat Bertotti-Robinson geometry $AdS_2\times S^2$ carries the Wu-Yang
	magnetic pole as an exact solution for arbitrary non-minimal couplings; the couplings
	only renormalise the radius of the $S^2$. The non-conformally-flat branch of solutions
	known for this system is recalled. The Hopf fibration $S^3/S^2\cong S^1$ of the gauge
	group manifold projects the Wu-Yang field on a uniform Dirac-type magnetic field, and the motion
	of a point electric charge in it is solved in closed form: the charge moves on a cone and
	oscillates between the horizon and a maximal radius.
\end{abstract}
\maketitle
	

\section{Introduction}\label{sec:intro}

The Bertotti-Robinson (BR) spacetime \cite{bertotti,robinson,dolan} is the product $AdS_2\times S^2$ of two-dimensional spaces of constant curvature with equal radii and opposite signs. It is the unique conformally flat, static, spherically symmetric solution of the source free Einstein-Maxwell equations with a nonzero electromagnetic field, and it describes a universe filled with a uniform, covariantly constant electromagnetic field. Its importance goes well beyond this: it is the near-horizon geometry of the extreme Reissner-Nordstr\"om black hole, and the symmetry enhancement $SL(2,\mathbb R)\times SO(3)$ of the throat is shared, in suitable form, by the near-horizon geometries of extremal black holes in general \cite{bardeen-horowitz,kunduri-lucietti}. The same $AdS_2\times S^2$ geometry underlies the attractor mechanism and the entropy function formalism for extremal black holes \cite{ferrara-kallosh-strominger,sen}, and $AdS_2$ gravity is a basic ingredient of attempts to understand extremal black hole microstates \cite{strominger}. More recently, BR has become the background in which exact black holes immersed in a uniform external field are constructed
\cite{podolsky-ovcharenko,ovcharenko-podolsky-2025,ovcharenko-podolsky-2026}.
Black holes in the universe do interact with external magnetic fields \cite{melvin,wald,ernst}, and such fields underlie the standard explanation of the high energy activity of
galactic nuclei and quasars \cite{blandford-znajek}. A uniform magnetic field on a black hole background is therefore a physically well motivated, and analytically tractable setting.

In the Einstein-Maxwell theory, the magnetic field of the BR universe is Abelian. It is natural to ask what happens when the gauge group is non-Abelian. The Einstein-Yang-Mills (EYM) system is known to be much richer than Einstein-Maxwell: it has particle-like solitons and colored black holes which have no analogue in the Abelian theory \cite{bartnik-mckinnon,volkov-galtsov}, and monopole type solutions with a finite-size horizon \cite{lee-nair-weinberg}. A basic non-Abelian magnetic configuration is the Wu-Yang monopole \cite{wu-yang} which is an $SU(2)$ gauge field of the hedgehog form that is, up to a singular gauge transformation, the Dirac monopole \cite{dirac1931} embedded in a $U(1)$ subgroup of $SU(2)$. It is related to the 't~Hooft-Polyakov monopole \cite{thooft,polyakov} as its singular, point like limit. In flat space the Wu-Yang monopole has an energy density $\propto 1/r^4$ and therefore a divergent total energy. In the BR geometry the situation is different. The radius of the $S^2$ is constant, the point $r=0$ is a horizon at infinite proper distance, and the energy density of the hedgehog is a finite constant, $|F|^2=1/r_0^4$. Thus the BR background provides a regular home for the Wu-Yang pole, which is one of the reasons why we study it here.

A second ingredient is the non-minimal coupling of gauge fields to curvature. Effective actions
obtained by integrating out massive fields on a curved background contain terms of the generic type $RF^2$. For QED they are the Drummond-Hathrell terms \cite{drummond-hathrell}, which modify the propagation of photons in curved spacetime; the general non-minimally coupled Einstein-Maxwell system with its Horndeski-type structure was studied in \cite{horndeski,balakin-lemos}, and the cosmological relevance of non-minimal Maxwell-curvature couplings in \cite{bamba-odintsov}. Couplings of this kind also arise from the dimensional reduction of higher-dimensional Euler form actions \cite{muller-hoissen,dereli-ucoluk}, which gives direct curvature Yang-Mills couplings for non-Abelian fields. For this reason a systematic study of the Yang-Mills version is of interest. In \cite{dereli-senikoglu} we derived, in the language of exterior differential forms, the field equations of an $SU(2)$ EYM theory
with the two independent non-minimal couplings $XR$ and $YR$, where $X=\hs\tr(\bm F\w\hs\bm F)$ is the Yang-Mills scalar and $Y=\hs\tr(\bm F\w\bm F)$ is the Pontryagin-type pseudoscalar, and found a particular class of Wu-Yang monopole solutions on a one parameter deformation of the BR metric. Non-minimal Wu-Yang monopoles were also considered in \cite{balakin-zayats}. In the present paper the Levi-Civita connection is enforced by a Lagrange-multiplier 2-form, so that co-frame, connection and gauge potential are varied independently and the field equations are obtained in a compact, coordinate-free form.

The aim of this paper is to complete that picture, and to use it. Our main results are the following. We keep the gauge coupling $g$ explicit, as the factor $1/2g^2$ in front of the Yang-Mills Lagrangian (it was set to one in \cite{dereli-senikoglu}), which allows us to identify the canonically normalised Abelian field and its magnetic charge $b=1/g$. Then we show that the exactly conformally flat BR geometry carries the Wu-Yang pole as an exact solution for arbitrary values of the non-minimal couplings $\gamma,\gamma'$ and for
vanishing cosmological constant. The couplings only renormalise the effective gravitational coupling and thereby the radius of the $S^2$, Eq.~(\ref{eq:r0}). The reason is that on BR the scalar curvature vanishes and the invariant $X$ is constant, so the non-minimal contributions to the field equations reduce to a rescaling of the Einstein tensor.
This case was not contained in \cite{dereli-senikoglu}. After that we recall the tuned, non-conformally flat branch of the earlier paper and give the relations among
$\gamma$, $\Lambda$ and the deformation parameter $k$. Using the Hopf fibration $S^3/S^2\cong S^1$ of the gauge group manifold \cite{trautman,ryder} we project the Wu-Yang field onto an Abelian one, which is a uniform Dirac-type magnetic field on $S^2$. We solve in closed form the motion of a point electric charge in this field \cite{poincare,kazama-yang-goldhaber}. The classical problem of a charge in the field of a monopole is integrable thanks to the conserved vector $\vec J$ which includes the Poincar\'e term; on the BR background the particle moves on a cone, and the radial motion is that of a harmonic oscillator in proper time, with a frequency that depends on the monopole charge. For $AdS_2\times S^2$ without a monopole the corresponding
geodesic problem has been discussed in \cite{alekseev2}.

The paper is organised as follows. In Sec.~\ref{sec:nonmin} we derive the non-minimally coupled EYM equations. Sec.~\ref{sec:WY} collects the properties of the BR metric and of its one-parameter family, and is devoted to the Wu-Yang solutions: the Yang-Mills sector, the exact solution on conformally flat BR, and the non conformally flat branch. In Sec.~\ref{sec:traj} we perform the Hopf projection and fix the magnetic charge, and solve the motion of a charged particle. Sec.~\ref{sec:concl}
contains a discussion and an outlook.

\section{Non-minimally Coupled Einstein-Yang-Mills Field Equations}\label{sec:nonmin}

Let $\bm A=A^jT_j$ be an $su(2)$-valued potential 1-form, $[T_j,T_k]=\epsilon_{jkl}T_l$, with field
strength $\bm F=\d\bm A+\bm A\w\bm A$, that is $F^j=\d A^j+\tfrac12\epsilon_{jkl}A^k\w A^l$.
The gauge coupling $g$ is not carried by the covariant derivative but multiplies the Yang-Mills action
(so $\bm A$ is $g$ times the canonically normalised potential). The Bianchi identity is
$\nabla_{\bm A}\bm F\equiv\d\bm F+\bm A\w\bm F-\bm F\w\bm A=0$, and the quadratic invariants are
\begin{equation}
	X=\hs\,\tr(\bm F\w\hs\bm F),\qquad Y=\hs\,\tr(\bm F\w\bm F),
\end{equation}
with $\tr(T_jT_k)=+\delta_{jk}$, so that $\tr(\bm F\w\hs\bm F)=F^j\w\hs F^j=|F|^2\hs1$. 
The Lagrangian density 4-form of Einstein-Yang-Mills theory is
\begin{equation}
	\mathcal L_0=\frac1{2\kappa^2}\,R\,\hs1+\frac1{2g^2}\tr(\bm F\w\hs\bm F),
\end{equation}
where the scalar curvature of spacetime is $R=-\hs(R_{ab}\w\hs e^{ab})$ and $\kappa^2=4\pi G$,
$G$ being Newton's universal constant. We add the non-minimal couplings, a cosmological term and the constraint term,
\ba
	\mathcal {L}&=&\mathcal L_0+\mathcal {L}_1+\Lambda\hs1+\mathcal L_C, \\
	\mathcal {L}_1&=&-\frac1{2g^2}\big(\gamma X+\gamma' Y\big)R\,\hs1,\nonumber \\
	\mathcal {L_C}&=&(\d e^a+\omega^a{}_b\w e^b)\w\lambda_a , \nonumber
\ea
where the Lagrange-multiplier 2-forms $\lambda_a$
impose the zero-torsion condition so that $\omega^a{}_b$ is the Levi-Civita connection.

Since $\hs\hs 1=-1$, one has $\tr(\bm F\w\hs\bm F)=-X\hs1$ and $\tr(\bm F\w\bm F)=-Y\hs1$, so that
\begin{equation}
	\mathcal L_0+\mathcal L_1=\frac{R}{2\kappa^2}\,\hs1-\frac1{2g^2}\Big(X+(\gamma X+\gamma' Y)R\Big)\hs1 .
\end{equation}
With $X=-|F|^2$ the Yang-Mills term therefore appears as $\frac1{2g^2}|F|^2(1+\gamma R)$, up to the $\gamma'$ contribution of $Y$.

Independent variations with respect to the co-frame and the potential give the equations:
\begin{align}
	&\Big(\frac1{\kappa^2}-\frac{\gamma X+\gamma' Y}{g^2}\Big)G_a-\Big(\Lambda+\frac{(\gamma X+\gamma' Y)R}{2g^2}\Big)\hs e_a \nonumber \\
	&=\frac1{g^2}\Big[(1+\gamma R)\,\tau_a[\bm F]-\gamma\,\D(\iota_a\hs\d X)-\gamma'\,\D(\iota_a\hs\d Y)\Big],\label{eq:Einstein}\\
	&\nabla_{\bm A}\bm F=0,\qquad \nabla_{\bm A}\hs\big(\bm F+\gamma R\bm F+\gamma' R\hs\bm F\big)=0,\label{eq:YM}
\end{align}
where $G_a=-\tfrac12R^{bc}\w\hs e_{abc}=G_{ab}\hs e^b$ are the Einstein 3-forms and
$\tau_a[\bm F]=\tfrac12\tr(\iota_a\bm F\w\hs\bm F-\bm F\w\iota_a\hs\bm F)=T_{ab}[\bm F]\hs e^b$
the Yang-Mills energy-momentum 3-forms. For $g=1$ these are eqs.~(12) and (15) of \cite{dereli-senikoglu}.

\section{Wu-Yang Magnetic Monopole Solutions in Bertotti-Robinson Spacetimes}\label{sec:WY}

\subsection{Bertotti-Robinson metric}

We give the metric, 
in isotropic coordinates $\{t,x,y,z\}$,
$r=\sqrt{x^2+y^2+z^2}$,
\begin{equation}
	\gm=-\Big(\frac r{r_0}\Big)^{2k}\d t^2+\Big(\frac{r_0}{r}\Big)^2(\d x^2+\d y^2+\d z^2), 	\label{eq:kmetric}
\end{equation}
which is $AdS_2\times S^2$ in topology for every real $k$ and conformally flat only for $k^2=1$.
Its curvature scalar is $R=-\frac2{r_0^2}(1-k^2)$, vanishing for $k^2=1$. In spherical polar coordinates $\{t,r,\theta,\varphi\}$ the metric reads
$$
\gm=-\Big(\frac r{r_0}\Big)^{2k}\d t^2+\Big(\frac{r_0}{r}\Big)^2 \d r^2 + {r_0}^2 ( d \theta^2 + \sin^2 \theta \d \varphi^2) . 
$$

\subsection{The Yang-Mills sector}
We consider the Wu-Yang potentials \cite{wu-yang}
\ba
	A^1&=&u\,(y\,\d z-z\,\d y),\quad A^2=u\,(z\,\d x-x\,\d z),\quad A^3=u\,(x\,\d y-y\,\d x), \nonumber \\
	u&=&\frac{K(r)-1}{r^2}.
\ea
Then
\begin{equation}
	F^1=\Big[(y^2+z^2)\frac{u'}{r}+2u+x^2u^2\Big]\d y\w\d z
	+\Big[\frac{u'}{r}-u^2\Big]x\,\d x\w(y\,\d z-z\,\d y),
\end{equation}
and $F^2,F^3$ follow by cyclic permutations of $(x,y,z)$.

\noindent Tracing over the Lie algebra, in the metric above,
\begin{equation}
	X=-\Big(\frac r{r_0}\Big)^4\Big(\frac{2K'^2}{r^2}+\frac{(K^2-1)^2}{r^4}\Big),\qquad Y=0 .
\end{equation}
For constant $K$ the Yang-Mills equation (\ref{eq:YM}) gives, on the metric (\ref{eq:kmetric}),
\begin{equation}
	\d\hs F^1+A^2\w\hs F^3-A^3\w\hs F^2
	=\Big(\frac r{r_0}\Big)^{k+1}\frac{K(K^2-1)}{r^3}\,\d t\w\d r\w\d x ,
\end{equation}
and cyclically for the other two. Because $R$ is constant for every $k$,
the non-minimal factor $(1+\gamma R)$ in (\ref{eq:YM}) does not change this. Hence $K=\pm1$
($\bm F=0$, the Yang-Mills vacua) or $K=0$, the non-trivial Wu-Yang pole. For $K=0$,
\begin{equation}
	F^j=-\frac{x^j}{r^4}\big(x\,\d y\w\d z+y\,\d z\w\d x+z\,\d x\w\d y\big)=-n^j \sin\theta\,\d\theta\w\d\varphi,\qquad
	n^j=\frac{x^j}{r},
	\label{eq:hedgehog}
\end{equation}
and
\begin{equation}
	X=-\frac1{r_0^4},\qquad Y=0,\qquad
	\frac1{2g^2}\tr(\bm F\w\hs\bm F)=\frac1{2g^2r_0^4}\,\hs1 .
\end{equation}

\subsection{Exactly conformally flat BR: $k=1$, $\Lambda=0$, arbitrary couplings}\label{sec:k1}

For $k=1$ the metric is the BR metric, with $R=0$, and $X$ is constant, so $\d X=0$ and $Y=0$.
Then Equation(\ref{eq:Einstein}) reduces to
\begin{equation}
	\Big(\frac1{\kappa^2}-\frac{\gamma X}{g^2}\Big)G_a-\Lambda\,\hs e_a=\frac1{g^2}\tau_a[\bm F].
\end{equation}
Both $G_a$ (since $R=0$) and $\tau_a$ (a Maxwell-type stress tensor) are traceless, so the trace
forces $\Lambda=0$. We get
\begin{equation}
	G_{ab}=\kappa_{\rm eff}^2\,T_{ab}[\mathcal F],\qquad
	\frac1{\kappa_{\rm eff}^2}=\frac1{\kappa^2}-\frac{\gamma X}{g^2}=\frac1{\kappa^2}+\frac{\gamma}{g^2r_0^4}.
\end{equation}
For $AdS_2\times S^2$ of equal radii, $G^\theta{}_\theta=\frac{1}{r_0^2}$ and $T^\theta{}_\theta=\frac{B^2}{2}$, so
\begin{equation}
	\frac1{\kappa^2}+\frac{\gamma}{g^2r_0^4}=\frac1{2g^2r_0^2}
	\quad\Longleftrightarrow\quad
	\frac{g^2}{\kappa^2}\,r_0^4-\frac12\,r_0^2+\gamma=0,
\end{equation}
that is,
\begin{equation}
	r_0^2=\frac{\kappa^2}{4g^2}\Big[1+\sqrt{1-\frac{16\,g^2\gamma}{\kappa^2}}\Big]
	\label{eq:r0}
\end{equation}
which is real for $\gamma\le\kappa^2/16g^2$ and for $0<\gamma\le\kappa^2/16g^2$ the root with the minus sign is also admissible. 

For $\gamma=0$ this is the Einstein-Yang-Mills BR radius $r_0^2=\kappa^2/(2g^2)$, i.e.\ that of an Einstein-Maxwell magnetic BR with charge $1/g$. The coupling $\gamma'$ does not appear since $Y=0$. The Yang-Mills equation (\ref{eq:YM}) holds since $R=0$ and $K=0$. Therefore
the Wu-Yang pole on the exactly conformally flat BR spacetime is a solution of the non-minimally
	coupled EYM system for any $\gamma,\gamma'$, with $\Lambda=0$ and the $S^2$ radius (\ref{eq:r0}).
This case is not contained in \cite{dereli-senikoglu}.

\subsection{The non-conformally flat branch}

For $\kappa^2\Lambda=1/(2\gamma)$ and $\gamma'=\gamma$ the Lagrangian factorises,
\begin{equation}
 \mathcal{L}_0+\mathcal{L}_1+\Lambda\hs1=\Big(\frac{R}{2\kappa^2}+\Lambda\Big)\Big(1-\frac{\gamma\kappa^2}{g^2}(X+Y)\Big)\hs1-\frac1{2g^2}\tr(\bm F\w\bm F),
\end{equation}
and the last term is topological (Pontryagin density), so it does not affect the field equations.
Since $X$ is constant and $Y=0$ on the Wu-Yang pole, both factors vanish, and so do all variations of the
product, provided
\ba
	1-\frac{\gamma\kappa^2}{g^2}X=0\ \Rightarrow\ \gamma=-\frac{g^2r_0^4}{\kappa^2},
\ea	
Leading to
	\ba
	\Lambda=\frac1{2\gamma\kappa^2}=-\frac1{2g^2r_0^4},\qquad
	R=-2\kappa^2\Lambda=\frac{\kappa^2}{g^2r_0^4}.
	\label{eq:conditions}
\ea
Combined with $R=-\frac2{r_0^2}(1-k^2)$ for the Bertotti-Robinson metric, the last relation fixes $k$:
\begin{equation}
	k^2=1+\frac{\kappa^2}{2g^2r_0^2}>1 .
\end{equation}
Relative to Eqs.~(35) and (36) of \cite{dereli-senikoglu}, the signs of $\gamma$ and $\Lambda$ are reversed and $k^2>1$ instead of $0<k^2<1$;
this follows from the sign of the Yang-Mills term in $\mathcal L_0$ and of $\mathcal{L}_1$ (positive energy for the Yang-Mills field, with $\gamma$ multiplying
$|F|^2R$ as in (\ref{eq:YM})).
This branch needs $\Lambda\neq0$ and tuned couplings; in contrast with the previous section
it is not conformally flat.


\section{Trajectories of a point electric charge}\label{sec:traj}


The gauge group manifold is $S^3\cong SU(2)$. The Hopf fibration $S^3/S^2\cong S^1$ singles out a $U(1)$
subgroup, which we identify with electromagnetic gauge group. The magnetic charge is counted by the degree of the Hopf map
\cite{trautman,ryder}. Projecting the $K=0$ field on the radial direction in isospace, with the canonical normalisation
$\mathcal F=\frac1g\,n^j F^j$, gives
\begin{equation}
	\mathcal F=-\frac1{g\,r^3}\big(x\,\d y\w\d z+y\,\d z\w\d x+z\,\d x\w\d y\big)=-b\,\sin\theta\,\d\theta\w\d\varphi,\qquad b=\frac1g,
\end{equation}
so that $\frac1{4\pi}\oint_{S^2}\mathcal F=-b$. In the gauge in which the potential vanishes at the South Pole ($\theta=\pi$),
\begin{equation}
	\mathcal A=b\,(1+\cos\theta)\,\d\varphi,\qquad \mathcal F=\d\mathcal A .
\end{equation}
For a probe of charge $q$ the Dirac condition is $q\,b\in\tfrac12\mathbb Z$.


On the exactly conformally flat BR background $R=0$, so the non-minimal factors in (\ref{eq:YM}) equal one and
the Abelian projection $\mathcal F$ obeys the free Maxwell equations; no dressing of $q$ or $b$ arises.
A probe of mass $m$ and charge $q$ follows
\begin{equation}
	m\,\hat\nabla_{\dot C}\dot C=-q\,\iota_{\dot C}\mathcal F,\qquad \gm(\dot C,\dot C)=-1,
\end{equation}
with a dot denoting $\d/\d\tau$ and $\tau$ the proper time; the sign follows from $m\hat\nabla_{\dot C}\dot C_\mu=q\mathcal F_{\mu\nu}\dot x^\nu$
and $(\iota_{\dot C}\mathcal F)_\nu=\dot x^\mu\mathcal F_{\mu\nu}$. 

In the Bertotti Robinson metric, the Lagrangian becomes
\begin{equation}
	\mathcal L=\frac m2\Big[-\frac{r^2}{r_0^2}\dot t^2+\frac{r_0^2}{r^2}\dot r^2+r_0^2(\dot\theta^2+\sin^2\theta\,\dot\varphi^2)\Big]+q\,b\,(1+\cos\theta)\dot\varphi ,
\end{equation}
and the cyclic coordinates $t,\varphi$ give the conserved quantities
\begin{equation}
	E=m\frac{r^2}{r_0^2}\dot t,\qquad
	p_\varphi=m r_0^2\sin^2\theta\,\dot\varphi+q\,b(1+\cos\theta).
\end{equation}
The vector $\vec J$ (orbital angular momentum plus the Poincar\'e term $-q\,b\,\hat r$ \cite{poincare}) is conserved,
so the charge moves on a cone of half-angle $\gamma_c$ with $\cos\gamma_c=-qb/J$. The angular part of the
normalisation condition is $r_0^2(\dot\theta^2+\sin^2\theta\,\dot\varphi^2)=(J^2-q^2b^2)/(m^2r_0^2)$, and the radial equation is
\begin{equation}
	\dot r^2+\frac{\mu^2}{r_0^2}\,r^2=\frac{E^2}{m^2},\qquad
	\mu^2=1+\frac{J^2-q^2b^2}{m^2r_0^2}\ \ge 1 .
\end{equation}
Using $\dot t=Er_0^2/(mr^2)$ we obtain
\begin{equation}
	r(\tau)=\frac{E r_0}{m\mu}\sin\Big(\frac{\mu}{r_0}\tau+\delta\Big),\qquad
	t-t_0=-\frac{m\mu r_0}{E}\cot\Big(\frac{\mu}{r_0}\tau+\delta\Big),
\end{equation}
and eliminating $\tau$,
\begin{equation}
	r(t)=\frac{E r_0/(m\mu)}{\sqrt{1+\Big[\dfrac{E}{m\mu r_0}(t-t_0)\Big]^2}} .
\end{equation}
The charge leaves the horizon $r=0$ as $t\to-\infty$, reaches
$r_{\max}=Er_0/(m\mu)$ at $t=t_0$, and falls back to the horizon as $t\to+\infty$,
keeping a constant inclination on the cone.

\begin{figure}[H]
	\centering
	\IfFileExists{rise.png}{\includegraphics[width=0.3\textwidth]{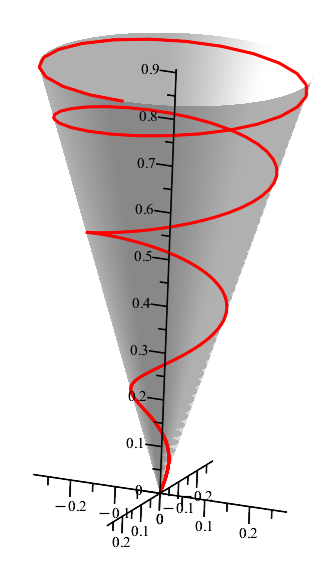}}{\fbox{\parbox{0.6\textwidth}{\centering Figure placeholder: ascent from the horizon}}}
	\caption{Trajectory showing the ascent from the horizon to the peak distance.}
\end{figure}

\begin{figure}[H]
	\centering
	\IfFileExists{fall.png}{\includegraphics[width=0.3\textwidth]{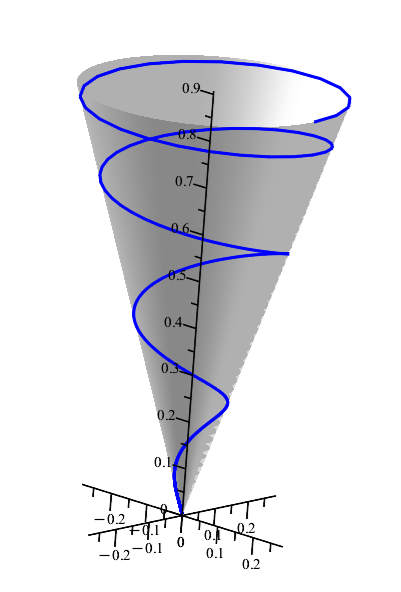}}{\fbox{\parbox{0.6\textwidth}{\centering Figure placeholder: descent to the horizon}}}
	\caption{Trajectory showing the descent from the peak distance back to the horizon.}
\end{figure}

\section{Conclusion and Outlook}\label{sec:concl}

We have studied the $SU(2)$ Einstein-Yang-Mills theory with non-minimal $XR$ and $YR$ couplings in the formalism of exterior forms, with the Levi-Civita connection imposed through a Lagrange multiplier, and with the gauge coupling $g$ kept explicit. Our main result is that the exactly conformally flat Bertotti-Robinson spacetime $AdS_2\times S^2$ carries the Wu-Yang magnetic pole for all values of the couplings $\gamma,\gamma'$,
with $\Lambda=0$. The mechanism is transparent. On BR the scalar curvature vanishes, so the factor $(1+\gamma R)$ in the Yang-Mills equation and in the stress tensor is unity, the pole has constant $X=-1/r_0^4$ and $Y=0$, so all derivative terms $\D(\iota_a\hs\d X)$ and $\D(\iota_a\hs\d Y)$ drop out, and the only surviving effect of the non-minimal terms is a shift of the coefficient of the Einstein tensor, $1/\kappa_{\rm eff}^2=1/\kappa^2+\gamma/(g^2r_0^4)$. The coupling $\gamma$ therefore acts as a renormalisation of Newton's constant by the Wu-Yang field itself, and it enters the geometry only through the radius of the $S^2$ Eq.~(\ref{eq:r0}). The coupling $\gamma'$ does not appear at all, because $Y=\hs\tr(\bm F\w\bm F)$ vanishes for a purely magnetic configuration. It can only matter when the gauge field has both electric and magnetic parts.

The radius formula (\ref{eq:r0}) has a few noteworthy features. The solution exists, with a real $r_0$, only for $\gamma\le\kappa^2/16g^2$; at $\gamma=0$ it reduces to the Einstein-Yang-Mills value $r_0^2=\kappa^2/2g^2$, that of an Einstein-Maxwell BR universe with magnetic charge $1/g$. For $0<\gamma\le\kappa^2/16g^2$ a second root exists, which at small $\gamma$ behaves as $r_0^2\simeq2\gamma$. Since $\gamma$ has dimension of length squared and the non-minimal terms are effective field theory corrections that are
controlled only when the curvature is small compared to $1/\gamma$, that is $r_0^2\gg\gamma$, the second root lies outside the regime where such a truncation can be trusted. Only the branch connected continuously to the Einstein-Yang-Mills solution at $\gamma\to0$ should be regarded as physical from the effective theory point of view. At $\gamma=\kappa^2/16g^2$ the two roots merge, $r_0^2=\kappa^2/4g^2$, and for larger $\gamma$ no BR solution of this type exists.

In contrast to this robust solution, the non conformally flat solutions on the deformed metric (\ref{eq:kmetric}) require tuned couplings, $\gamma'=\gamma$, $\kappa^2\Lambda=1/2\gamma$, $\gamma=-g^2r_0^4/\kappa^2$, and a non-zero cosmological
constant, and they fix the deformation parameter to $k^2=1+\kappa^2/2g^2r_0^2$. We have corrected the signs of $\gamma$, $\Lambda$ and the range of $k$ with respect to \cite{dereli-senikoglu}, so that the Yang-Mills term has positive energy and
$\gamma$ multiplies $|F|^2R$ as in the Yang-Mills equation (\ref{eq:YM}). The resulting $k^2>1$ branch is a genuinely different geometry from the BR solution, which is the $k=1$ member of the family.

Through the Hopf fibration of the gauge group manifold the pole becomes a uniform Dirac-type
magnetic field on $S^2$ with canonically normalised charge $b=1/g$, and the Dirac condition $qb\in\frac12\mathbb Z$ then quantises the ratio $q/g$ of the probe charge to the gauge coupling. The motion of a charged particle in this field is completely integrable. With the conserved energy $E$ and the conserved vector $\vec J$ (orbital angular momentum plus the Poincar\'e term), the charge moves on a cone of half-angle $\gamma_c$ with
$\cos\gamma_c=-qb/J$, and the radial motion is a harmonic oscillation in proper time,
$r(\tau)=(Er_0/m\mu)\sin(\mu\tau/r_0+\delta)$, with frequency $\mu/r_0$ and
$\mu^2=1+(J^2-q^2b^2)/m^2r_0^2\ge1$. Thus the $AdS_2$ throat acts as a confining trap, and the effect of the monopole charge and of the angular momentum is to stiffen it. The particle leaves the horizon, reaches $r_{\max}=Er_0/m\mu$, and returns to the horizon after a finite proper time $\pi r_0/\mu$ but an infinite coordinate time,
$r(t)\to0$ as $t\to\pm\infty$, as expected for a Poincar\'e-type horizon. The probe is a test particle and the motion does not depend on $\gamma,\gamma'$ except through the radius $r_0$, because $R=0$ and the Abelian projection satisfies the free Maxwell equations.

Two limitations should be stated. The BR solution is only the near-horizon geometry: we have not constructed an asymptotically flat solution of the non-minimally coupled equations that interpolates between flat space and BR, so it is open whether the Wu-Yang hair survives away from the throat. We have also not studied the stability of the solution under non-spherical or non-Abelian perturbations, which is a natural question since many EYM solutions are unstable \cite{volkov-galtsov}.

We see several directions for further work. The Ricci-type couplings $R_{ab}F^{ab}\w\hs F$ and $Ric_a\w\iota^aF\w\hs F$, which do not vanish on BR although the Weyl couplings do, should modify the BR solution non-trivially therefore they should be included to determine how universal the renormalisation of the $S^2$ radius is. Dyonic configurations, with electric and magnetic Yang-Mills fields, have $Y\neq0$ and are the natural place where $\gamma'$ becomes visible. The entropy-function approach \cite{sen} applied to the
non-minimal Lagrangian should give the corrections of the horizon entropy of the extreme black hole with Wu-Yang magnetic charge. The full black-hole solution interpolating between the flat asymptotic region and the BR throat, the
extreme Reissner-Nordstr\"om black hole with Wu-Yang charge in this theory, could be sought numerically. The quantum mechanics of charged scalar and spinor particles in the
$AdS_2\times S^2$ background with a monopole, based on monopole harmonics \cite{wu-yang-76} and on the Dirac equation in the field of a monopole \cite{kazama-yang-goldhaber}, will lead to a spectrum whose angular part is controlled by $qb\in\frac12\mathbb Z$, and to a radial problem on $AdS_2$ that is closely related to the classical oscillation found here. Other gauge groups and other non-Abelian embeddings of the Hopf construction could be treated by the same method. Finally, the presence of a uniform, magnetically charged BR region in a non-Abelian theory suggests that magnetic fields around astrophysical black holes may carry non-Abelian structure in a sufficiently strong-curvature regime, which would be of interest for the astrophysical applications.

\section*{Acknowledgement}
We dedicate this work to Prof. Metin G\"urses for his 60 Years in Mathematical Physics. One of us (T.D.) thanks the Turkish Academy of Sciences (T\"UBA) for partial support.
\section*{Data Availability Statement}
No new data were created or analyzed in this study.

\end{document}